# Elastic models of the fast traps of carnivorous *Dionaea* and *Aldrovanda*

Marc Joyeux[#]

*Laboratoire Interdisciplinaire de Physique (CNRS UMR 5588),*

*Université Joseph Fourier Grenoble 1, BP 87, 38402 St Martin d'Hères, France*

**Abstract**: The carnivorous aquatic Waterwheel Plant (*Aldrovanda vesiculosa* L.) and the closely related terrestrial Venus Flytrap (*Dionaea muscipula* SOL. EX J. ELLIS) both feature elaborate snap-traps, which shut after reception of an external mechanical stimulus by prey animals. Although *Aldrovanda* is usually considered as a miniature aquatic *Dionaea*, the shutting mechanisms of the two plants are actually quite different. The fast shutting of *Aldrovanda* is indeed based on a simple swelling/shrinking mechanism, while the movement of *Dionaea*'s traps is accelerated by the snap-buckling of the lobes. The purpose of this Report is to describe several key improvements to the elastic models that have recently been introduced to elucidate these movements [Poppinga and Joyeux, *Phys. Rev. E* 84, 041928 (2011)]. In particular, a precise mechanism for the action of the motor cells of *Aldrovanda* is proposed, the facts that the opening of the leaves of *Dionaea* is an irreversible mechanism based on growth and that the strain field is anisotropic and much smaller on the inner than on the outer surface of the leaves during shutting are taken properly into account, and a more accurate formula for calculating mean curvatures is used. The improvements brought to the model are described in detail and the physical consequences of these improvements are discussed.

[#] email adress : Marc.Joyeux@ujf-grenoble.fr

# I - Introduction

Carnivorous plants are an active field of research from both the botanical [1] and biophysical [2] points of view. Of particular interest are the fast traps of some species, which are able to capture insects within time scales as short as a few milliseconds [3,4]. It has been suggested that the physical mechanisms at play in these movements actually depend on the size of the organism, the physical environment (*e.g.* the density of the medium), and the characteristic times that must be achieved [5]. Repetitive and reversible bending movements are usually produced by changes in volume and length of certain specialized cells (called *motor cells*) relative to antagonistic motor cells or to neighboring non-motor cells of fixed volume [6]. Such changes are driven by the turgor pressure variations that result from the transport of $K^+$ or $Cl^-$ ions through the cell membrane. Faster repetitive movements usually rely on the rapid geometric changes of specific organs due to the snap-buckling of thin membranes [7]. Since it is often difficult to determine with certainty from the mere analysis of high-speed video recordings which mechanism actuates the rapid plant movement that is being observed, we recently proposed models based on the theory of elastic solid thin plates to help characterize the physics involved in such movements [4,8,9]. Using these models, we were in particular able to show that, although *Aldrovanda* is usually considered as a miniature aquatic *Dionaea*, these two carnivorous plants actually display different shutting mechanisms. The fast shutting of *Aldrovanda* is indeed based on a simple swelling/shrinking mechanism, while the movement of *Dionaea*'s traps is accelerated by the snap-buckling of the lobes [8].

The models presented in Ref. [8] can however be improved with respect to several points. Concerning *Aldrovanda*, it was for example shown in Ref. [8] that the opening/closing movements of the traps can be actuated by decreasing/increasing the curvature of the midrib, but no mechanism accounting for the curvature changes was proposed. A more detailed description of the trap mechanism is therefore desirable. Moreover, it was assumed that the closed conformation corresponds to a minimum of the elastic energy, which implies that the elastic force that drives the shutting of the lobes decreases strongly during the movement. This is not very realistic, because the plant would have to supply actively and rapidly quite a lot of energy from another source in order to overcome the drag force exerted by water and finish closing the trap. It was similarly assumed in Ref. [8] that the minimum energy conformation of *Dionaea* is the closed geometry. This is again disputable, because it is known that the opening of *Dionaea*'s traps is an irreversible mechanism based on growth [10] and that spontaneous curvatures correspond to the open geometry [7,11]. This indicates that the conformation with minimum elastic energy is actually the open geometry. Moreover, it was assumed in Ref. [8] that the strain field is isotropic, while experiments suggest that the snapping mechanism of *Dionaea* is due to an anisotropic elongation of the outer layer perpendicular to the midrib (see Fig. 2 in Ref. [7]). At last, it was assumed in Ref. [8] that mean curvatures are positive quantities, so that curvature energy is of the same order of magnitude for convex and concave leaf geometries, which is probably not reasonable. This point may be important in the case of *Dionaea*, for which snapping is accompanied by a very rapid inversion of the curvature of the lobes [7]. The sign of mean curvatures should therefore be taken properly into account in the model.

The purpose of this Report is to show how the models introduced in Ref. [8] can be adapted to take these remarks into account and to discuss the physical consequences of these improvements. The elastic model is described briefly in section II, in order to highlight the changes made with respect to Ref. [8]. The trapping mechanisms of *Dionaea* and *Aldrovanda* are next discussed in sections III and IV, respectively.

**II - The elastic model**

The elastic energy of solid thin plates can be written as the sum of a strain (in-plane) contribution and a curvature (out-of-plane) contribution, $E_{pot} = E_{strain} + E_{curv}$, where [12,13]

$$E_{strain} = \frac{Eh}{2(1-\nu^2)} \int_S [(1-\nu)\text{Tr}(\boldsymbol{\varepsilon}^2) + \nu(\text{Tr}(\boldsymbol{\varepsilon}))^2] \, dS$$
$$E_{curv} = \frac{Eh^3}{24(1-\nu^2)} \int_S [(\text{Tr}(\mathbf{b}))^2 - 2(1-\nu)\text{Det}(\mathbf{b})] \, dS . \quad (2.1)$$

In this equation, $E$ stands for the Young's modulus of elasticity of the leaf, $h$ for its thickness, $\nu$ for its Poisson's ratio, $\mathbf{b}$ for the difference between the strained and unstrained local curvature tensors, $\boldsymbol{\varepsilon}$ for the two-dimensional Cauchy-Green local strain tensor, and integration is performed over the surface $S$ of the leaf. In the simulations presented thereafter, it was assumed that $\nu = 0.50$, as for potato tubers [14,15] (note that force-extension results depend very little on the exact value of $\nu$ in the range 0.3 to 0.5 [16]), and a typical value $E = 5$ MPa [16-20] was used for the Young's modulus. For numerical purposes, the traps were described as triangle meshes with $M$ triangles (or facets) and $N \approx M/2$ vertices. The curvature energy was estimated according to

$$E_{curv} = \frac{Eh^3}{24(1-\nu^2)} \sum_{j=1}^{N} [(c_{j1} + c_{j2} - c_{j1}^0 - c_{j2}^0)^2 - 2(1-\nu)(c_{j1} - c_{j1}^0)(c_{j2} - c_{j2}^0)]\delta A_j, \quad (2.2)$$

where $\delta A_j$ denotes the elementary area associated to vertex $j$ (Eq. (3.2) of Ref. [4]), $c_{j1}$ and $c_{j2}$ the principal curvatures at vertex $j$ for the current geometry, and $c_{j1}^0$ and $c_{j2}^0$ the principal curvatures for the reference geometry (*i.e.* the spontaneous curvatures). The product $c_{j1}c_{j2}$ was estimated according to Eq. (3.12) of Ref. [4], while the sum $c_{j1} + c_{j2}$ was obtained from

$$c_{j1} + c_{j2} = \frac{1}{2\delta A_j} \sum_{k \in N_1(j)} (\cot \alpha_{jk} + \cot \beta_{jk})[\mathbf{n}_j \cdot (\mathbf{r}_k - \mathbf{r}_j)], \quad (2.3)$$

where $\mathbf{n}_j$ denotes the normal to the surface directed outwards at vertex $j$ (Eqs. (3.14) and (3.9) of Ref. [4]). Note that the values of $c_{j1}+c_{j2}$ obtained from Eq. (2.3) can be either positive or negative, while Eq. (3.9) of Ref. [4], which was used in previous work [3,4,8], leads to values of $c_{j1}+c_{j2}$ that are always positive. The values of $c_{j1}$ and $c_{j2}$ were finally deduced from $c_{j1}+c_{j2}$ and $c_{j1}c_{j2}$ using standard relations.

As in Ref. [8], two mechanically coupled meshes were built from the original one, in order to model the several cell layers that compose the vegetal tissue. The inner and outer meshes have respective thickness $h^{\text{in}}$ and $h^{\text{out}}$ satisfying $h = h^{\text{in}} + h^{\text{out}}$ and are computed according to $\mathbf{r}_j^{\text{in}} = \mathbf{r}_j - \tfrac{1}{2}h^{\text{in}} \mathbf{n}_j$ and $\mathbf{r}_j^{\text{out}} = \mathbf{r}_j + \tfrac{1}{2}h^{\text{out}} \mathbf{n}_j$, where $\mathbf{r}_j$ denotes the position vector of vertex $j$ of the original mesh. The total strain energy is the sum of the contributions of the two meshes, $E_{\text{strain}} = E_{\text{strain}}^{\text{in}} + E_{\text{strain}}^{\text{out}}$, where each contribution was estimated according to

$$E_{\text{strain}}^{\Lambda} = \frac{E\,h^{\Lambda}}{2(1-\nu^2)} \sum_{n=1}^{M} [(1-\nu)\text{Tr}((\boldsymbol{\varepsilon}_n^{\Lambda})^2) + \nu(\text{Tr}(\boldsymbol{\varepsilon}_n^{\Lambda}))^2]\,\delta S_n^{\Lambda}\;. \qquad (2.4)$$

In this equation, $\Lambda$ stands for "in" or "out" and $\delta S_n^{\Lambda}$ denotes the elementary area of facet $n$ of mesh $\Lambda$. The differential variations of turgor pressure inside each layer were modeled by assuming that they modify the separations between the vertices of the reference geometry according to a given function $\mathbf{S}$. These modified reference separations are used to compute the Gram matrices

$$\mathbf{F}_n^{0,\Lambda} = \begin{pmatrix} \mathbf{S}(\mathbf{r}_{n2}^{0,\Lambda}-\mathbf{r}_{n1}^{0,\Lambda})\cdot\mathbf{S}(\mathbf{r}_{n2}^{0,\Lambda}-\mathbf{r}_{n1}^{0,\Lambda}) & \mathbf{S}(\mathbf{r}_{n2}^{0,\Lambda}-\mathbf{r}_{n1}^{0,\Lambda})\cdot\mathbf{S}(\mathbf{r}_{n3}^{0,\Lambda}-\mathbf{r}_{n1}^{0,\Lambda}) \\ \mathbf{S}(\mathbf{r}_{n2}^{0,\Lambda}-\mathbf{r}_{n1}^{0,\Lambda})\cdot\mathbf{S}(\mathbf{r}_{n3}^{0,\Lambda}-\mathbf{r}_{n1}^{0,\Lambda}) & \mathbf{S}(\mathbf{r}_{n3}^{0,\Lambda}-\mathbf{r}_{n1}^{0,\Lambda})\cdot\mathbf{S}(\mathbf{r}_{n3}^{0,\Lambda}-\mathbf{r}_{n1}^{0,\Lambda}) \end{pmatrix}, \qquad (2.5)$$

where $\mathbf{r}_{n1}$, $\mathbf{r}_{n2}$ and $\mathbf{r}_{n3}$ denote the positions of the three vertices of facet $n$, and the Cauchy-Green local strain tensors $\boldsymbol{\varepsilon}_n^{\Lambda}$ according to Eq. (2.3) of Ref. [8]. While linear functions $\mathbf{S}(\mathbf{r}) = \sqrt{1 \mp \tfrac{1}{2}s}\,\mathbf{r}$, where $s$ is the varying strain parameter, were used in Ref. [8], the relevance of using different functions $\mathbf{S}$ will be discussed shortly.

Energy minimization was performed as described in Ref. [8]. Time evolution of *Dionaea's* traps during snapping was computed by solving numerically Langevin equations with internal damping with a leapfrog algorithm (Eq. (2.8) of Ref. [8], with $\gamma$=300 ms$^{-1}$). Langevin equations with external damping were instead considered for *Aldrovanda*, in order to take external water into account (Eq. (3.13) of Ref. [4], without the pressure and thermal noise terms, and with $\gamma$=30 ms$^{-1}$).

**III – The snap-buckling mechanism of *Dionaea***

The starting point of the new model is the parametric surface with $N$=1073 vertices and $M$=2048 triangles defined in Eq. (4.1) of Ref. [8], except that parameter $D$ was set to 4.5 mm instead of 3.0 mm to make the trap somewhat broader. As in Ref. [8], the leaf thickness was assumed to be $h$=400 μm with, however, $h^{in}$=300 μm and $h^{out}$=100 μm. The trap was first opened by imposing to the leaves a strain field defined by $\mathbf{S}(\mathbf{r}) = \mathbf{r}$ (for the inner mesh) or $\mathbf{S}(\mathbf{r}) = \sqrt{1+s}\,\mathbf{r}$ (for the outer mesh) for values of *s* decreasing from 0 to -0.11. In agreement with the remark made in the Introduction, *the obtained conformation*, which is labeled A in Fig. 1, *was taken as the reference conformation for all subsequent calculations*. Shutting of the trap was actuated by imposing to the open leaves (conformation A) a strain field close to the experimentally observed one reported in Fig. 2 of Ref. [7]. The measured strain field is indeed oriented almost perpendicular to the midrib and decreases farther from it. This was mimicked by using a function $\mathbf{S}(\mathbf{r}) = \mathbf{r}$ for the inner mesh and a function $\mathbf{S}(\mathbf{r}) = \sqrt{1+s(1-\beta)}\,\mathbf{r}_{//} + \sqrt{1+s/5}\,\mathbf{r}_{\perp}$ for the outer mesh, where $\alpha$ and $\beta$ are the two parameters that describe the leaf shape (see Eq. (4.1) of Ref. [8]) and $\mathbf{r}_{//}$ and $\mathbf{r}_{\perp}$ are the components of **r** parallel and perpendicular to the plane $x = z\tan\alpha$, respectively. The value of the strain parameter *s* was increased from 0 to 0.60. As can be seen in the bottom plot of Fig. 1, which shows the evolution with *s* of the elastic energy of the trap, the system evolves smoothly for values of *s* up to 0.40 (conformation B). At *s*=0.40, the leaves buckle from the almost-open conformation B to the closed conformation C. Note that a contact term must be added to the energy function to prevent the two leaves from interpenetrating (conformation C*).

The model proposed here differs from that of Ref. [8] with respect to several points, namely (i) the conformation with minimum elastic energy is the open geometry, (ii) the active role is played by the sole outer layer, (iii) the strain field is highly anisotropic, and (iv) the sign of mean curvatures is properly taken into account. Nonetheless, Fig. 1 confirms that shutting of *Dionaea*'s traps is greatly accelerated by the snap-buckling of the leaves, which corresponds to a jump of the system from an energy curve associated with essentially convex geometries (line AB) to an energy curve associated with essentially concave geometries (line CD). Moreover, the plots of the time evolution of the lobe aperture and average curvatures, which are shown in Fig. S1 [21] for a variation rate of *s* of 1.0 s$^{-1}$ (see Movie S1 [21]), compare well with experimentally measured ones [7,9] and with those obtained with the former model [8]. The main difference between the two models is that the new one predicts that only a small amount of energy is dissipated during buckling (that is, the energy of conformation C is only slightly smaller than that of B), while the former model predicted a large amount of energy dissipation (see Fig. 5 of Ref. [8]). The reason for this difference is, of course, that the former model assumed that the closed geometry has minimum elastic energy.

**IV – The swelling/shrinking mechanism of *Aldrovanda***

The starting point of the new model is the parametric surface with $N$=2113 vertices and $M$=4096 triangles defined in Eq. (3.1) of Ref. [8]. The leaf thickness was assumed to be $h$=70 μm, with $h^{in} = h^{out}$ =35 μm. The main question consists in understanding how narrow stripes of motor cells surrounding the midrib can open and close the leaves [22]. Such stripes were modeled as the regions of the parametric surface that satisfy $0 \leq \beta \leq \beta_{max}$, where $\beta$ is the parametric angle in Eq. (3.1) of Ref. [8]. The value $\beta_{max} = 18°$ was chosen so as to get narrow bands, as illustrated in insets A, B and C of Fig. 2, where the region such that $0 \leq \beta \leq \beta_{max}$ is highlighted in red. It was found that the only way to open and close the trap efficiently and realistically consists in imposing an anisotropic strain field that acts only on the inner layer and parallel to the midrib (*i.e.* in the *xz* plane [8]), according to $\mathbf{S}(\mathbf{r}) = \mathbf{r}$ for the outer mesh and $\mathbf{S}(\mathbf{r}) = \sqrt{1 + s f(\beta)}\, \mathbf{r}_{//} + \mathbf{r}_{\perp}$ for the outer mesh, where $\mathbf{r}_{//}$ and $\mathbf{r}_{\perp}$ are the components of $\mathbf{r}$ parallel and perpendicular to the plane $y = 0$, respectively, and $f(\beta)$ is a cut-off function that was introduced in order to avoid numerical problems that arise when neighboring facets have too different mechanical properties. $f(\beta)$ is defined according to $f(\beta) = 5[\beta / \beta_{max}]^6 - 6[\beta / \beta_{max}]^5 + 1$ for $0° \leq \beta \leq \beta_{max}$ and $f(\beta) = 0$ for $\beta > \beta_{max}$. $f(\beta)$ decreases smoothly from 1 at $\beta = 0$ (*i.e.* on the midrib) to 0 at $\beta = \beta_{max}$, with horizontal tangents at 0 and $\beta_{max}$. The high exponents insure that the interval where $f(\beta)$ is close to 1 is sufficiently broad. The fact that the strain field is imposed only to the inner layer is consistent with the experimental observation that the active motor cells are the inner epidermal cells of the motor zone [22] and that the rapid shutting in *Aldrovanda* is accompanied by a marked migration of K$^+$ and Cl$^-$ ions in the trap lumen (see Ref. [23] and Ref. [10], pp. 104-105). The strain field was first imposed to the initial parametric surface for values of *s* decreasing from 0 to -0.10 and *the obtained conformation*, which is labeled C in Fig. 2, *was taken as the reference conformation for all subsequent calculations*. The trap could subsequently be opened and closed by submitting conformation C to the same strain field for values of $s$ varying between 0.10 (closed conformation, labeled B in Fig. 2) and 0.60 (open conformation, labeled A in Fig. 2). As can be seen in the main plot of Fig. 2, which displays the evolution with $s$ of the elastic energy of the trap, no snap-buckling occurs during these movements and the system just goes back and forth along a single energy curve. The fact that conformation C is used instead of B as the reference conformation insures that the elastic force does not decrease too much during shutting, as discussed in the Introduction. A contact term must again be added to the energy function to prevent the two leaves from interpenetrating (as in conformation C) when $s$ is decreased below 0.10. The plot of the time evolution of the lobe aperture, which is shown in Fig. S2 [21] for a variation rate of $s$ of -5 s$^{-1}$ (see Movie S2 [21]), looks quite similar to the experimentally measured one [9], while average curvatures evolve only little during shutting, in sharp contrast with the case of *Dionaea*.

The present model therefore confirms that narrow stripes of motor cells located along the midrib are indeed sufficient to actuate the movements of *Aldrovanda*'s traps. There is, however, one important difference between the present model and that of Ref. [8]. Indeed, the strain field described above tends to bend the midrib region inwards (respectively, outwards) during opening (respectively, shutting). This is illustrated in Fig. 2, where the insets labeled [A] and [B] show the conformations of the stripes with $0 \leq \beta \leq \beta_{max}$ when the remainder of the leaves is discarded. Comparison of these insets with insets A and B indicates that, under the influence of the remainder of the leaves, the midrib actually deforms in the opposite directions compared to [A] and [B]. Indeed, the midrib bends outwards during opening and inwards during shutting [10,22]. The conclusion drawn in Ref. [8], according to which curvature variations along the midrib drive the trap movements, was therefore too simplistic. A more correct conclusion is rather that trap movements are actually driven by the mechanical coupling between the strain field imposed by the motor cell stripes and the remainder of the leaves.

# FIGURE CAPTIONS

**Figure 1** (color online) : Variations of $E_{pot}$, $E_{strain}$, and $E_{curv}$ as a function of *s* (bottom) and calculated *Dionaea*'s trap conformations at various points of the diagram (top). Solid and dashed lines describe two different energy surfaces, which are associated with convex and concave leaf geometries, respectively. Conformation C* is the buckled leaf geometry when contact terms preventing leaf interpenetration are omitted. Conformation D is the limiting concave leaf geometry that becomes unstable upon decrease of *s* and transforms to a convex geometry through a buckling event similar to the one leading from B to C.

**Figure 2** (color online) : The main plot shows the variations of $E_{pot}$, $E_{strain}$, and $E_{curv}$ as a function of *s*. Insets labeled A, B and C show the calculated conformations of *Aldrovanda*'s trap at various points of the diagram. Insets labeled [A] and [B] show the conformations at point A and B of the stripes with $0 \leq \beta \leq \beta_{max}$ when the remainder of the leaves is discarded.

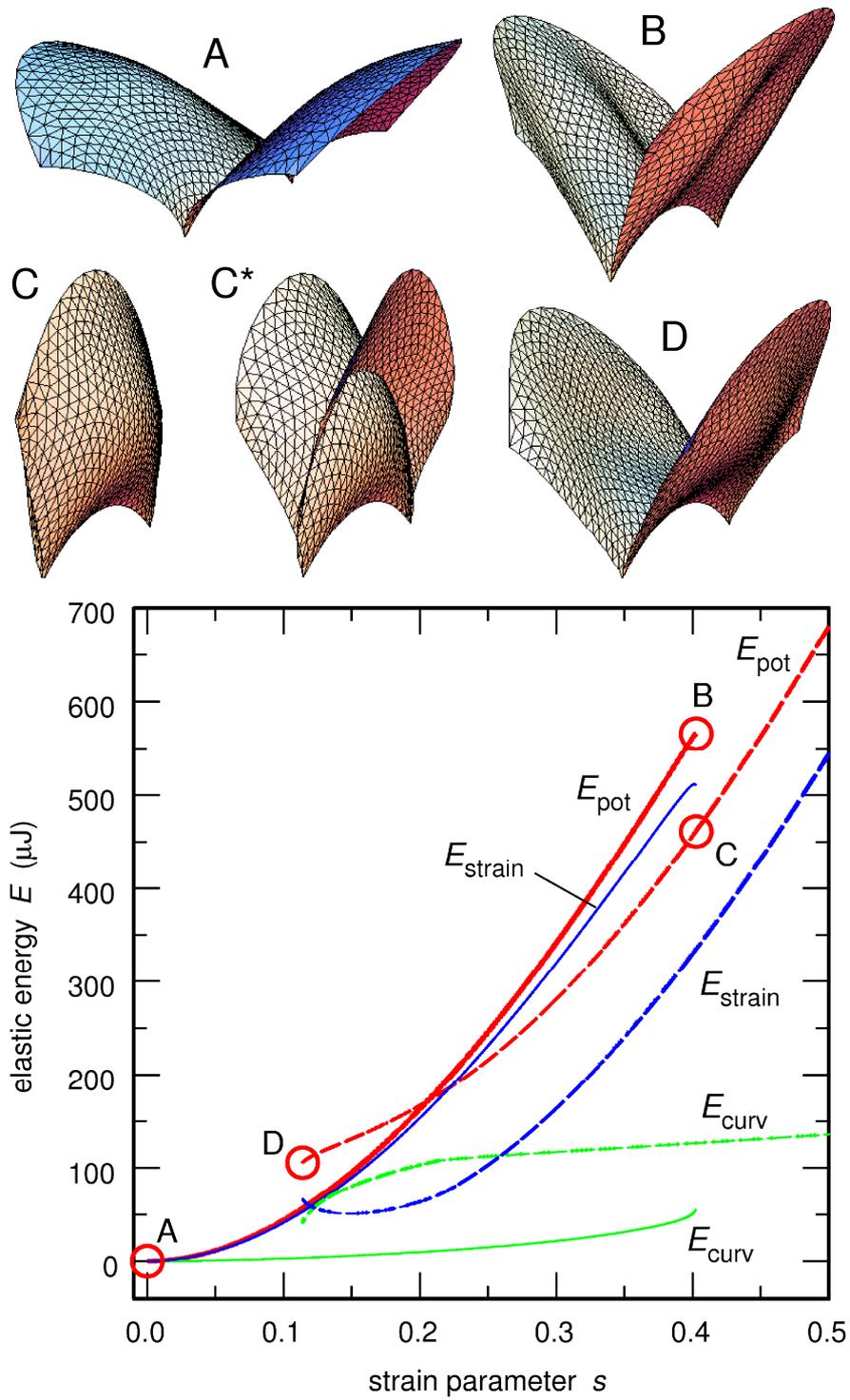

**FIGURE 2**

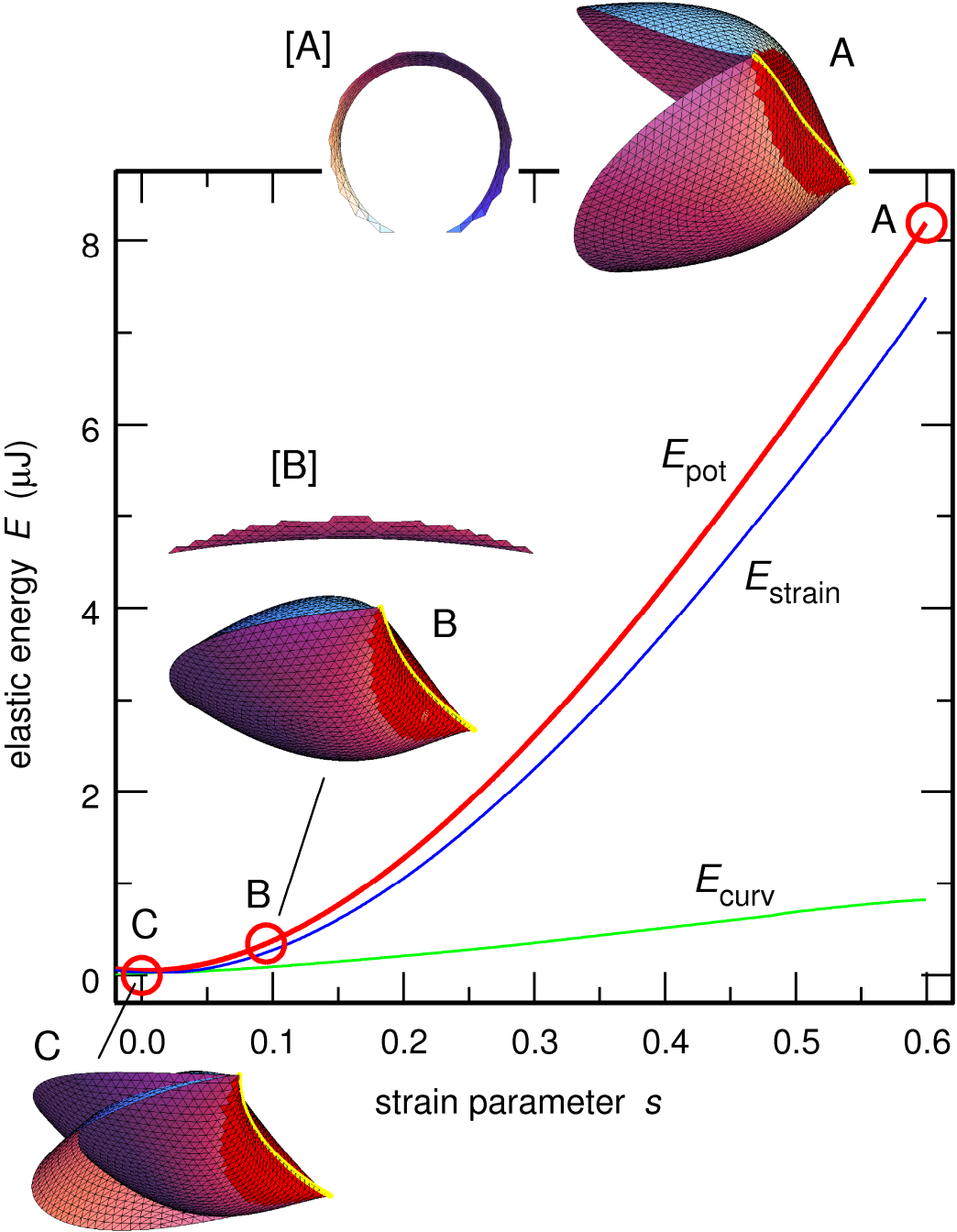

**FIGURE S1** (.jpg, 570 Ko) : Plot of the time evolution of the lobe aperture (top plot) and average mean and Gaussian curvatures (bottom plot) of a *Dionaea*'s trap during the snapping event shown in Movie S1 of the Supplemental Material.

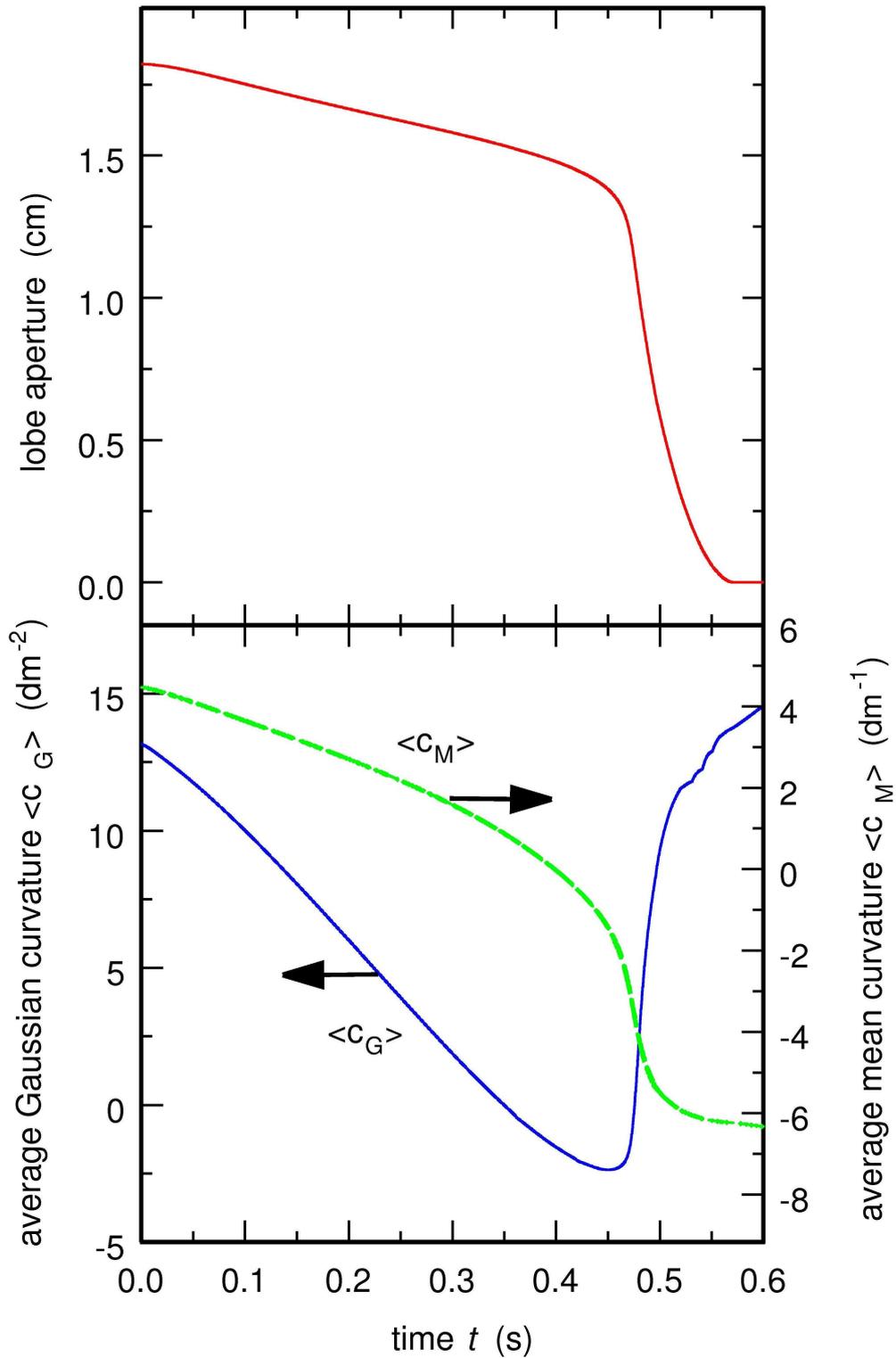

**FIGURE S2** (.jpg, 540 Ko): Time evolution of the lobe aperture (top plot) and average mean and Gaussian curvatures (bottom plot) of an *Aldrovanda*'s trap during the shutting event shown in Movie S2 of the Supplemental Material.

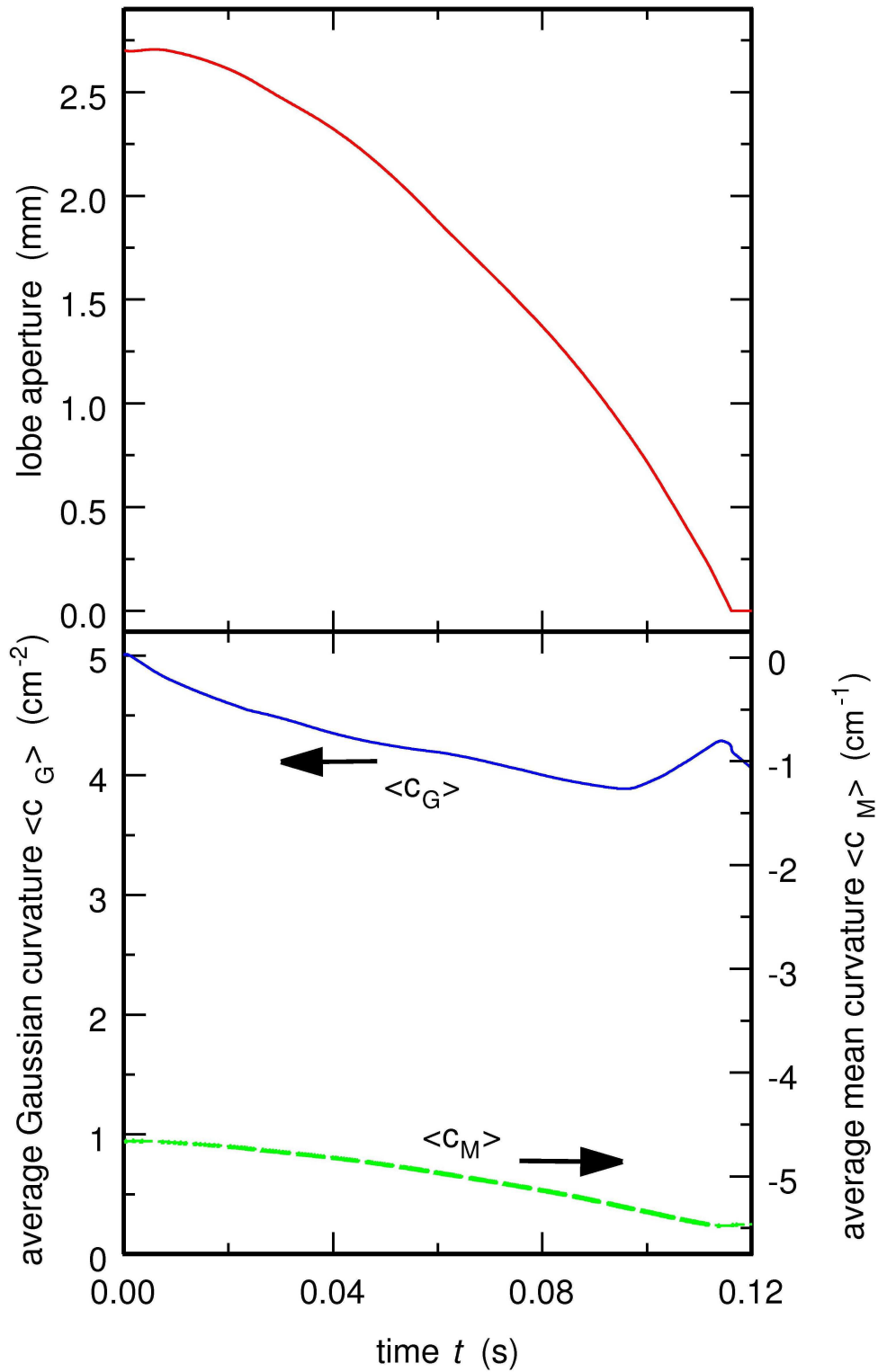